# Hybrid Simulation-based Resource Planning and Constructability analysis of RCC Pavement Projects

Shakerian, Mohammad[1], Rajabi, Mohammad Sadra[2], Tajik, Mohammad[1] and Taghaddos, Hosein[3*]

[1] M.Sc. Student, School of Civil Engineering, College of Engineering, University of Tehran, Tehran, Iran
[2] Ph.D. Student, Department of Civil and Environmental Engineering, The George Washington University, Washington DC, United States
[3] Associate Professor, School of Civil Engineering, College of Engineering, University of Tehran, Tehran, Iran
* corresponding author: htaghaddos@ut.ac.ir

**Abstract:** One of the critical challenges in infrastructural constructions is designing and planning operations and their related resources. The complex interlinked composition of different factors and variables affecting resource productivity has made simulation a powerful approach for operational planning. The construction sector has recently seen a notable surge in applying various simulation tools to enhance further the quality of projects' planning, particularly in large-scale infrastructure developments (e.g., highway construction). Due to possible cost overruns in improper resource allocation, optimizing the design and construction planning stages of megaprojects such as massive pavement projects is essential. Recent studies aimed to build a simulation-based strategy in construction designing and planning by combining various simulation approaches (e.g., discrete-event simulation, system dynamics, agent-based simulation, and hybrid simulation) to enhance the planning phase. This paper introduces an evolving real-time hybrid simulation technique regarding the project's intrinsic time-varying inputs and factors to optimize the planning of Roller Compacted Concrete (RCC) pavement projects. Several scenarios are investigated using various resource combinations to achieve the best execution method for delivering concrete to the project. An actual highway project case study validates the proposed model and its application for future projects. This study's findings exhibit the proficiencies of the simulation-based approach in resource planning of RCC pavement projects within the time and cost constraints and their related regulations.

## 1    INTRODUCTION

efficient scheduling and planning phase of a construction project significantly impact its likelihood of success. However, such a significant step requires a considerable amount of work and resources due to the complexity of construction projects. A better understanding of construction operations is obtainable through visual modeling of construction activities.

Traditional planning and scheduling methods encounter many shortcomings when applied to complex construction projects. Pritsker (1989) has presented some of these inadequacies. For instance, conventional planning approaches are limited due to their static and deterministic assumptions. Traditional techniques are more suitable for projects with fewer activities and fewer uncertainties. Critical Path Method (CPM), Line Of Balance (LOB), and Program Evaluation and Review Techniques (PERT) are a few of these mentioned traditional scheduling methods (Abdallah & Marzouk, 2013).

Therefore, simulation modeling has been introduced to develop computer-based representations of construction systems to recognize their behavior (AbouRizk, 2010; AbouRizk et al., 2011). In previous studies, mathematical modeling systems such as mathematical programming and queuing theory did not demonstrate suitable applications in large-scale and more complex construction projects. To address these issues, computer simulation techniques were introduced as an effective tool to design and analyze construction procedures despite their complexity or size. Computer models can generate the proper solutions while considering the overall logic of various activities, the resources involved, and the environment under which the project is constructed (Ekyalimpa et al., 2012). Simulation models are entirely successful in embodying the process of building a facility and can be used to develop better strategies, optimize resource utilization, minimize costs or project duration, and improve overall construction project management (AbouRizk, 2010).

Numerous simulation tools have been introduced during recent years in which CYCLONE, STROBOSCOPE, Simphony.Net, and VitaScope are among the most recent ones (Hajjar & AbouRizk, 1999; Halpin, 1977). Sawhney et al. (1998) describe the enhancements made to the CYCLONE modeling methodology to allow simultaneous simulation of processes involved in a construction project. These enhancements aim to develop individual components for all the processes that constitute a project and then link them to simulate them simultaneously using a shared resource pool. Although simulation modeling is focused on the planning stages, better decision-making requires modeling the project during the construction and planning phases (Dra et al., 2016).

Recent studies have employed Discrete Event Simulation (DES), Agent-based simulation, system dynamics (i.e., continuous simulation), or hybrid simulation integrating discrete event simulation and system dynamics models to model construction environments. Such studies model a construction project from different perspectives, such as effective resource allocation, safety enhancement, site layout planning (Taghaddos et al., 2012, Alzraiee et al. 2015, (Goh & Askar Ali, 2016, Taghaddos et al., 2021). In hybrid simulation,  feedback loops play a major role to automate the process (Salloum et al., 2020).

The focus of this study is to employ hybrid simulation to enhance the supply chain management of RCC pavement projects.  The most relevant study for this paper is (Taghaddos & Dashti, 2017), which introduces a simulation-based approach to model a significant milestone of the construction process in an RCC dam. Several what-if construction scenarios were analyzed for the milestone based on different combinations of the resources, and then a time-cost trade-off analysis was used to find the optimum construction scenario for the milestone. The proposed approach gives the managers a vision of the total time and cost of the project based on the different alternative scenarios.

However, the construction of RCC dams and RCC pavement have significant differences. RCC pavement is amongst the linear project, making supply chain management (i.e., concrete supply from batching plant to the paver) more challenging. The following chapter provides a brief background to RCC pavement projects.

## 1.1    RCC Pavements

Roller Compacted Concrete (RCC) is a durable, economical, and low-maintenance material for low-speed heavy-duty paving applications, including industrial and transit pavements. Recently, there has been increasing interest in the use of RCC for public roadways. Its consistency is rigid enough that it can maintain its stability under the load of the roller compactor with vibration and, at the same time, allow a proper spreading of the concrete mixture without any separation. Similar to many other construction projects, pavement construction involves complex geometry and complicated processes (Delatte, 2004). RCC heavy-duty pavements may be constructed in single or multiple layers depending on the design capacity of the road (Piggot, 1986). Generally, roller compacted concrete is spread by a modified asphalt paver and compacted by steel drum roller, vibrating roller, and rubber wheel roller (Chhorn et al., 2017).

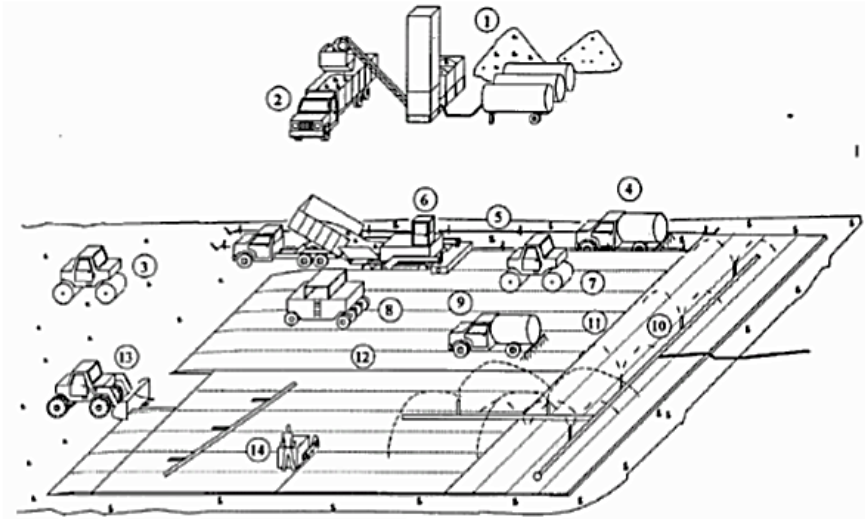

Figure 1: Schematic illustration of RCC pavement construction activities. Captured from Iranian guideline for the construction of concrete roads (issue #733)

Figure 1 shows a visual sequence of activities that conclude an RCC pavement project. Dump trucks are usually used to haul the concrete mix to the paving front. Iranian concrete pavement guideline allows for a maximum elapsed time of 45 minutes after mixing has occurred. This factor is most dominant when the pavement site or work zone is far from the batching plant. Compaction of the placed RCC must follow immediately after its placement (Piggot, 1986). Curing and cutting wedges are another two primary operations.

A steady concrete supply to the paver machine must be maintained for civil code compliance and better surface quality. Planning of hauling machines gets more complicated when project geometry is constantly changing. In road projects, the relative distance of resources is essential in decision-making. Also, few constraints for maintaining the quality of concrete affect planning and resource scheduling.

This paper proposes a simulation model to optimize resources for concrete supply to the paver using different capacity dump trucks while meeting all the constraints issued by concrete-based road guidelines. For better results, this study embeds a dynamic feedback system to allow automated alteration of resources regarding the state and progress of the project.

## 2.    Overview of Case Study (Babak-Herat RCC Highway 2nd Lane)

Babak city is located in Tehran-Bandar Abbas transit road, and Babak-Herat transit road sits at the heart of Iran's east-west transport corridor. It is located on the Sassanid historic road and was built by the National Construction and Development of Transport Infrastructure Organization in 2016. The second lane of this highway with a length of 44 kilometers and width of 11 meters is the subject of this study, used to develop an accurate model for future planning of the same type of projects. In order to obtain a high-quality surface in the Babak-Herat project, according to the employer's recommendation, a paving machine with an 11 meters width distribution capability was rented. Unlike conventional concrete paving, roller concrete paving is made without molds, dowels, or steel reinforcements. It is usually unnecessary to make a joint with a saw, but if required by the technical specifications, the distances of transverse joints are longer than concrete paving. The concrete plant sits at a distance of 25 kilometers from the start point (Babak city) in the middle of the proposed road.

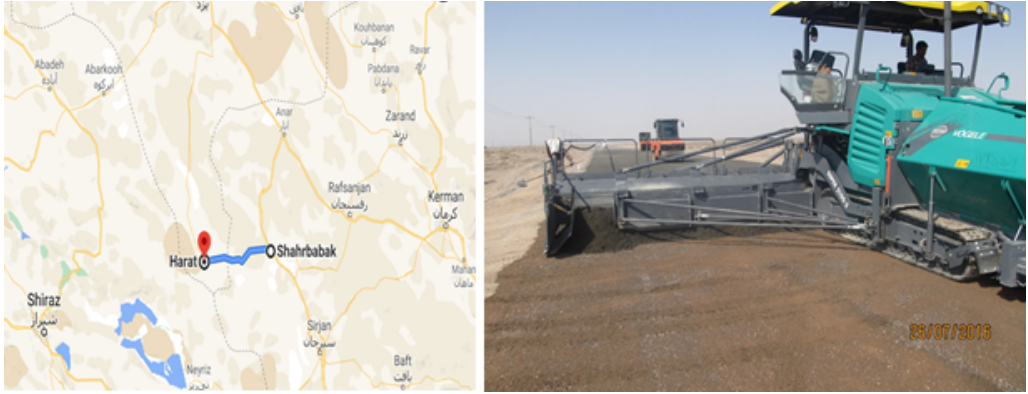

Figure 2: Location of project and paving machine used on the project

## 3.    Proposed Multi-Step Methodology

### 3.1    Modeling of the RCC Pavement Construction

As shown in figure 3, a linear RCC pavement project is modeled using Simphony.Net 4.6. Generally, linear constructions are defined as the projects in which most of the work consists of highly repetitive activities. These operations are repeated in each location for the entire road length. The flow of concrete from batching plant through dump trucks and into the paver must be modeled continually to control regulations and guidelines more effectively. Therefore, a combination of DES and continuous simulation has been applied. This model is also equipped with a system dynamic method with the same general premise of Figure 4.

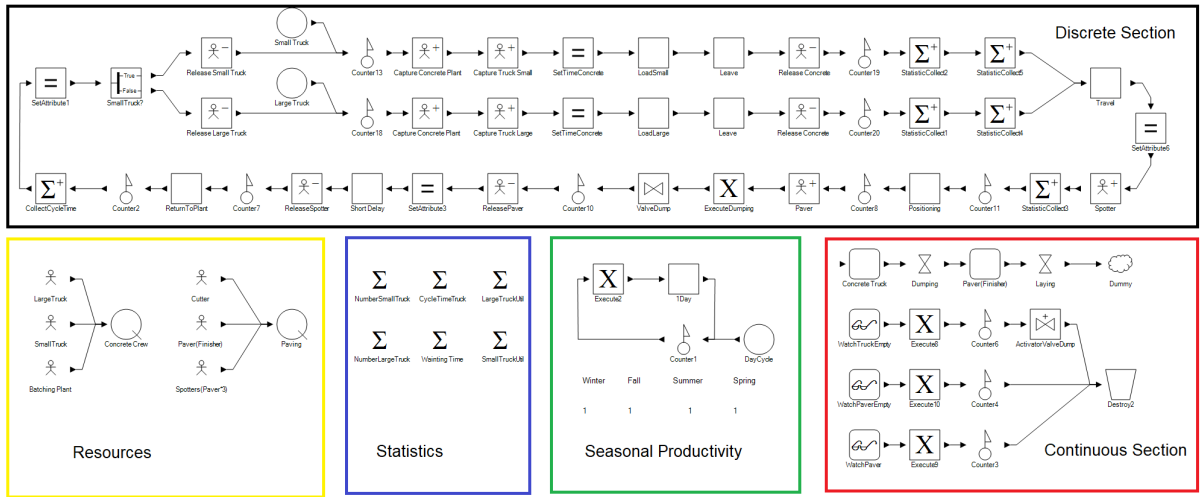

Figure 3: A developed model for RCC pavement construction in Simphony .Net

Due to the nature of road projects, equipment location and their relative distance are in constant change. Hence behavior of activities follows functions related to the progress of simulation and state of other operations. This model uses four feedback loops to allocate resources better as the project advances as its relative distances and durations change. This feature can help the process to adapt to the available resources at a given time. For example, similar to this case study, if the concrete plant is at the middle point of the road, as the project progresses, the pavement frontline gets closer to batching plant, and a smaller fleet of trucks is required to maintain a steady flow of concrete to the paver. After passing the middle point of the road, the relative distance to the concrete plant increases, and the number of required resources changes.

The proposed model in this study focuses on the supply chain of concrete to the paver, as these operations are affected the most by the project's geometry. For example, the number of required trucks would be much lower in the middle point of the construction rather than the start point since the work zone moves toward the batching plant. All project inputs, durations, and relative distances are user-defined. This ability enhances the model's customizability; therefore, it can be used for similar projects such as asphalt pavements without much effort in changing its components and inner properties. This study considers all the essential tasks in an actual RCC pavement project which correlate to the concrete supply resources.

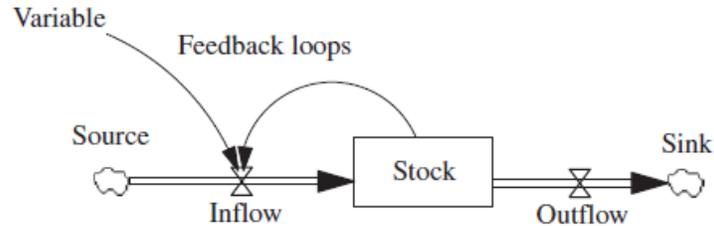

Figure 4: A simple diagram of a system dynamic model. Adapted from (Sterman, 2000)

## 3.2    Scenario Analysis for Various Combinations of Resources

Pavement construction is composed of many resources linked and reliant on each other. Table 1 shows the main activities in an RCC pavement project, its duration, and necessary resources. One of the primary resources and the object of this study is the dump trucks used to convey concrete to the paver on site. Dump trucks with different capacities are often used concurrently in the same project. Large dump trucks can carry 7.5 cubic meters of concrete in this study, 50% more than small dump trucks. The main object of this scenario analysis is to find the optimum number of large and small dump trucks while satisfying the government's RCC regulations and those imposed by the owner. Two main constraints in this model are listed below.

1.    Concrete must be placed and compacted within 45 minutes.

2.    For a steady concreting process, dump trucks must maintain an interarrival time of less than 3 minutes.

All possible combinations must be modeled to find an optimum scenario with the highest utilization rate. These different scenarios must be filtered to satisfy the owner's constraints and mandatory regulations. The maximum number of large and small dump trucks is 10 and 5, respectively. In theory, 50 different scenarios must be checked to find the optimum dump truck combination for this project. Table 2 shows the accepted composition of dump trucks and the mean utilization of each type in every scenario. This table indicates that scenario #39 is the optimum resource grouping since it has a higher average utilization. Also, due to the limited capacity of paver for laying concrete, increasing the number of dump trucks after a certain point will only result in higher waiting times and long queues.

Table 1: Main activities and their required resources

| Activity* | Duration (minutes) | Resources |
|---|---|---|
| Loading small truck | 3 | Small truck, Batching plant |
| Loading large truck | 4.5 | Large truck, Batching plant |
| Travel to paver | Inconstant* | Small or Large truck |
| Dumping | 2.5(small truck) 3.75(large truck) | Small or Large Truck, Paver, Spotters |
| Return to batching plant | Inconstant* | Small or Large truck |

*Travel distances are continuously changing during the project

Table 2: Accepted scenarios for dump trucks

| Scenario Number | Number of Large Dump Trucks | Number of Small Dump Trucks | Mean Utilization Large Dump Truck (%) | Mean Utilization Small Dump Truck (%) | Total Project Duration (minutes) |
|---|---|---|---|---|---|
| #39 | 9 | 4 | 65.5 | 7.7 | 96919 |
| #40 | 10 | 4 | 58.9 | 7.7 | 96919 |
| #49 | 9 | 5 | 65.5 | 6.1 | 96919 |
| #50 | 10 | 5 | 58.9 | 6.1 | 96919 |

## 3.3 Enhancement of Resource Planning with System Dynamics

After finding the optimum scenario, this model is integrated with a system dynamic approach to obtain more detailed resource planning while the project progresses. Since road projects are constantly changing location, the number of required resources might need to update along the way. Feedback loops can change the number of resources while the simulation progresses. As the pavement advances, the frontline of work is getting closer to the batching plant in this study. Therefore, the number of dump trucks could be updated to maintain a high utilization.

Figure 5 shows the utilization of large dump trucks without any feedback loops during pavement construction based on the optimum scenario determined in the last step. The average utilization rate of large dump trucks is 65.5%. After finding the optimum resource combination for the dump trucks, the SD method, explained in methodology, is enabled for detailed planning of the resources throughout the project's lifespan. Figure 6 shows the utilization of large dump trucks being constantly updated using feedback loops during the construction. The average utilization rate with this enhanced method is 89.9%. This figure also shows the effectiveness of the system dynamics method for yielding a more significant average utilization rate.

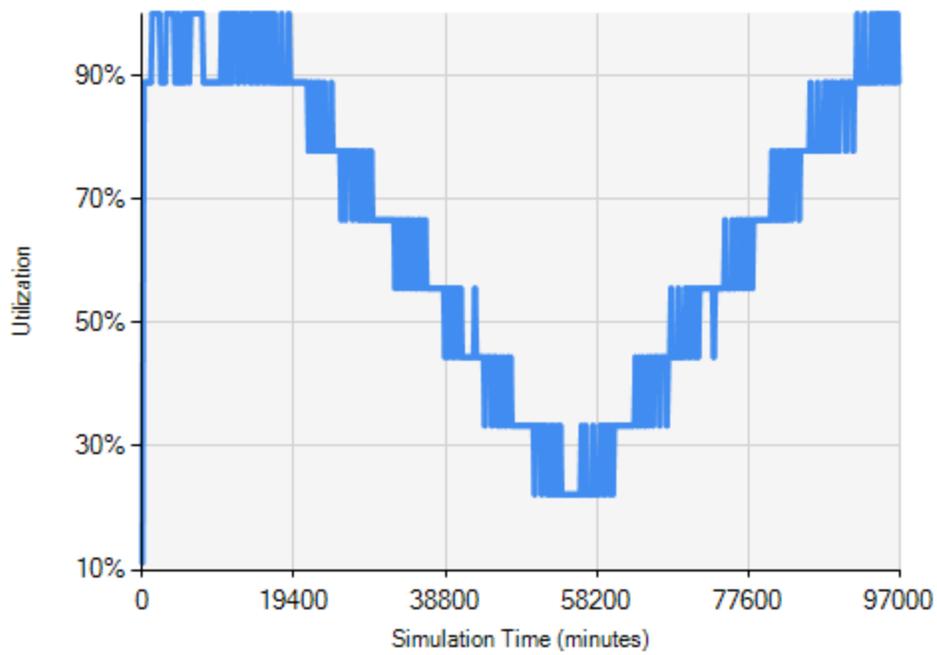

Figure 5: Utilization rate of large dump trucks

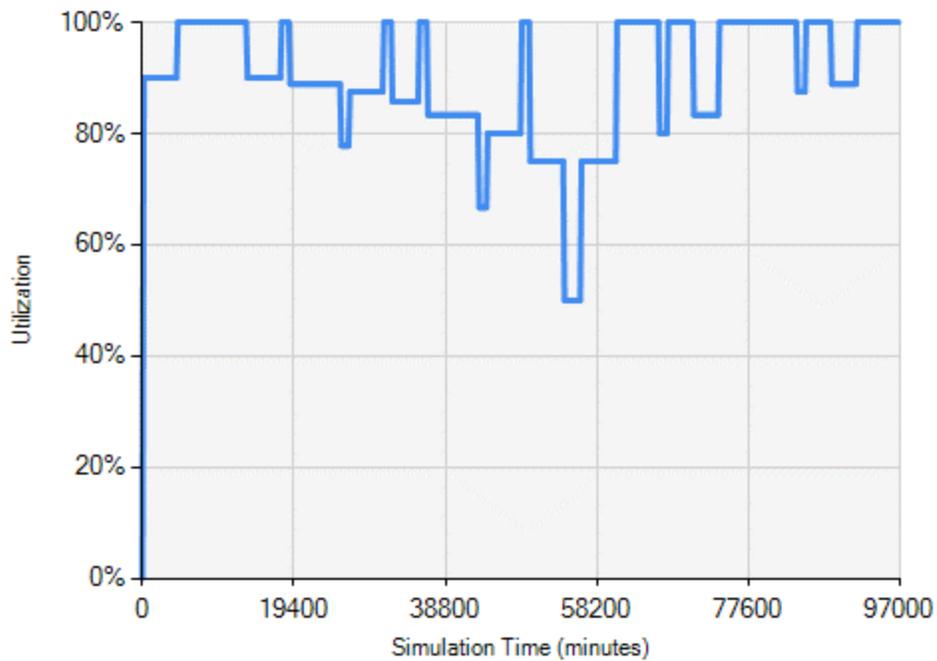

Figure 6: Enhanced utilization rate of large dump trucks

Figure 7 shows detailed resource planning as this study's main result and output for Babak-Herat RCC Road. A detailed resource allocation plan is presented for construction managers and industry practitioners.

With this method, engineers will be able to reallocate these excess resources to other projects to reduce the total cost of projects.

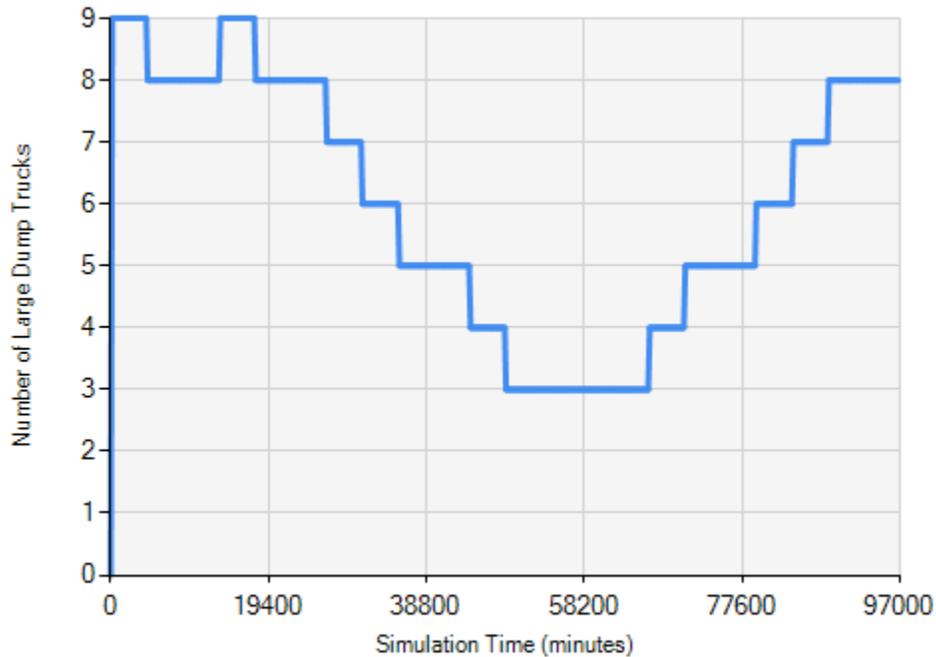

Figure 7: Final enhanced resource scheduling of large dump trucks

## 4. LIMITATIONS

The proposed model in this study is limited to RCC pavement but can be effortlessly extended to include other pavement materials such as hot mix asphalt (HMA). Optimization of resources unrelated to work zone geometry has been neglected for simplicity. Since these resources do not correlate with any variables in their behaviors, static and traditional planning methods based on experts' opinion was opted instead. The main reason behind all the above limitations is that this study only includes the specific optimization and level of detail required by the project owner.

## 5. CONCLUSION

This study acknowledges the importance and advantages of simulation-based modeling in construction projects' planning and scheduling phase. Due to construction projects' dynamic and complicated characteristics, simulation tools have been utilized to model construction processes during recent years. Simulation-based modeling is a powerful tool capable of visualizing and modeling construction activities. Road, pavement, and resurfacing projects are critical infrastructural developments due to their enormous budget and importance to the public. Therefore, having an optimum resource and execution plan which meets all the governmental policies ensures lower cost and better product quality.

This paper proposes a dynamic hybrid simulation-based approach to find the optimum scenario based on the available resources and constraints in an actual RCC pavement construction project. It utilizes mathematical equations based on the continuously changing durations in a horizontal project to regulate and adapt to the new geometry.

The primary purpose of this study was to find the optimum resource combination of different dump truck types that are mainly involved in the RCC supply chain to the paver at the frontline. All the possible combinations were tested concerning the government guidelines on concrete pavement construction. Optimum resource combination is then selected based upon higher utilization of resources. A case study of the RCC pavement project is also investigated using a hybrid model in Simphony .Net. The results illustrate the capabilities of the simulation-based approach in designing and planning pavement construction projects.

To validate the proposed model's results, the optimum scenario was compared to those used by the contractor in Harat-Shahrbabak highway $2^{nd}$ lane in 2016. The comparison showed a better utilization rate but close to the actual project. Hence, this dynamic model can be used in future projects for better resource allocation and planning.

The overall results of this study demonstrate the proficiencies of simulation tools in planning resources of pavement projects. Simulation models can be programmed to develop more detailed resource schedules, yielding higher utilizations.

## 6.    ACKNOWLEDGEMENTS

The authors would like to gratefully acknowledge Mr. Khazaeli, Tecnosa R&D center at the University of Tehran, and Abad-Garan Inc. for their kind collaboration and assistance.